\documentclass[12pt]{iopart}
\jl{1}
\usepackage{psfig}
\begin{document}
%%%%%%%%%%%%%%%%%%%%%%%%%%%%%%%%%%%%%%%%%%%%%%%%%%%%%%%%%%%%%%%%%%%%%%%%%%%%%%%
\def\<{\langle}
\def\>{\rangle}
\def\({\left (}
\def\){\right )}
\def\[{\left [}
\def\]{\right ]}
\def\i{{\rm i}}
\def\e{{\rm e}}
\def\d{{\rm d}}
\def\/{\over}
\def\s{\mbox{s}}
%%%%%%%%%%%%%%%%%%%%%%%%%%%%%%%%%%%%%%%%%%%%%%%%%%%%%%%%%%%%%%%%%%%%%%%%%%%%%%
\title[Reaction matrix for Dirichlet billiards]
{Reaction matrix for Dirichlet billiards with attached waveguides}
\author{Holger Schanz\footnote{holger@chaos.gwdg.de}} 
\address{Max-Planck-Institut f\"ur Str\"omungsforschung und Institut
   f{\"u}r Nichtlineare Dynamik der Universit{\"a}t G{\"o}ttingen,
   Bunsenstra{\ss}e 10, D-37073 G\"ottingen} 
\date{\today}
\begin{abstract}
  The reaction matrix of a cavity with attached waveguides connects scattering
  properties to properties of a corresponding closed billiard for which the
  waveguides are cut off by straight walls. On the one hand this matrix is
  directly related to the S-matrix, on the other hand it can be expressed by a
  spectral sum over all eigenfunctions of the closed system. However, in the
  physically relevant situation where these eigenfunctions vanish on the
  impenetrable boundaries of the closed billiard, the spectral sum for the
  reaction matrix, as it was used before, fails to converge and does not
  reliably reproduce the scattering properties. We derive here a convergent
  representation of the reaction matrix in terms of eigenmodes satisfying
  Dirichlet boundary conditions and demonstrate its validity in the
  rectangular and the Sinai billiards.
\end{abstract}
\date{\today}
%\maketitle
%%%%%%%%%%%%%%%%%%%%%%%%%%%%%%%%%%%%%%%%%%%%%%%%%%%%%%%%%%%%%%%%%%%%%%%%%%%%%%%
Recently, there has been some interest in the application of the
reaction-matrix theory of Wigner and Eisenbud \cite{WE47} and the
projection-operator formalism of Feshbach \cite{Feshbach}, originally
developed for the description of nuclear collisions, to chaotic cavities with
attached scattering channels \cite{Dit00,PSS01,AR01,S+02,SSS02}. Such models
are frequently used as paradigms of chaotic scattering
\cite{DS92a,B+92,SS95,HKL00} and found important experimental realizations by
electron transport through open quantum dots \cite{M+92,S+98}, lasing optical
micro-cavities \cite{G+98b,C+99,H+01b} and scattering of microwaves in
resonators with attached waveguides \cite{P+00,S+02}. It is known that the
quantum scattering in the open system shows signatures of the classical
dynamics in the closed system. For example, conductance fluctuations of open
quantum dots are different for systems whose closed counterparts have
integrable, fully chaotic or mixed phase space \cite{M+92,S+98}.  In fact many
statistical results for quantum chaotic scattering rely on this connection as
for their derivation an ad-hoc formulation of the scattering problem in terms
of an effective non-Hermitian Hamiltonian is used \cite{Feshbach,MW69}, which
describes the dynamics inside a closed system and an additional coupling of
the corresponding eigenstates to some scattering channels. Classical chaos
enters via the random-matrix assumption for the Hermitian part of this
Hamiltonian \cite{VWZ85,FS97,GMW98}. However, while the effective Hamiltonian
appears naturally within the formalism from nuclear physics, it is not a
priori clear when this formalism applies to some given billiard system and how
the parameters of the two different models are related.

Therefore it is important and interesting to develop a thorough understanding
of the connection between the scattering properties of billiards, in
particular the S-matrix, and the properties of the corresponding closed
system, i.~e.\ spectrum and eigenfunctions. For example, today it is well
known that the spectrum can be found from a secular equation involving the
S-matrix \cite{DS92a,B+92,SS95}.  For scattering from the outside of convex
billiards in ${\cal R}^{2}$ this scattering approach to quantization allows
even for a mathematically rigorous formulation \cite{EP97}.

Unfortunately, the opposite direction is less profoundly understood. Here, the
unitary S-matrix is related to a Hermitian reaction matrix, and this can in
turn be expressed as a sum over the internal spectrum with coefficients
reflecting the behavior of the internal wavefunction at the boundary
separating billiard and wave\-guide. From the physical point of view, and in
particular for the aforementioned applications
\cite{M+92,S+98,G+98b,C+99,H+01b,P+00}, the most natural situation is a
wavefunction which vanishes outside the billiard and on the boundary
(Dirichlet b.~c.): Electrons in a quantum dot are depleted from the boundary
by a high negative gate voltage, the radiation field is restricted to the
optical cavity by total internal reflection or additional mirrors, and in
microwave resonators the metallic walls do not admit the electrical field.
However, for this of all choices of boundary conditions no consistent
representation of the reaction matrix as a spectral sum is known.  We derived
a formal expression \cite{Dit00,PSS01} but found both numerically and from
semiclassical estimates that it fails to converge.  Recently, this conclusion
was confirmed with different methods \cite{SSS02}.

It is certainly possible to circumvent this problem: One option is to ignore
the divergence and to restrict the spectral sum by hand to some set of levels.
Naturally this must fail in a generic situation, but it can give reasonable
results in some special cases, e.~g.\ when the spectrum has a doublet or band
structure \cite{SSS02}. A second option relies on the fact that in principle
all boundary conditions providing a self-adjoint Hamiltonian are admissible
for defining the closed system, see \cite{SSS02} for a nice explanation of
this point. In particular Neumann b.~c.~with finite wave function but zero
derivative at the interface were considered previously
\cite{Dit00,PSS01,AR01,S+02,SSS02}. In this case the spectral sum converges,
and the resulting reaction matrix successfully reproduces numerical or
experimental scattering data.  Nevertheless it would be very unsatisfying,
could the formulation of a proper reaction-matrix theory for billiards not be
based on the physically relevant boundary conditions.

To remedy this situation we establish in the present paper a convergent
expansion of the reaction matrix for Dirichlet billiards with attached
waveguides. Our main result is Eq.~(\ref{W}) below, where the coupling
constants entering this series are given. Before we get to this equation, we
recall some results from previous work. Following Eq.~(\ref{W}) we give a
derivation of this formula, discuss some interesting aspects related to it and
show with two examples how it works.

%%%%%%%%%%%%%%%%%%%%%%%%%%%%%%%%%%%%%%%%%%%%%%%%%%%%%%%%%%%%%%%%%%%%%%%%%%%%%%%

\begin{figure}[tb]
  \centerline{\psfig{figure=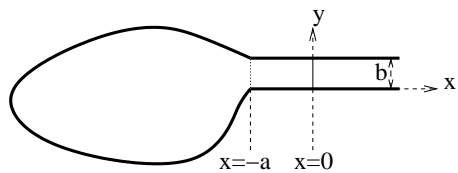,width=60mm}}
  \caption{\label{sketch} A scattering system consisting of a cavity
    and an attached waveguide of width $b$ is shown with bold lines. The
    coordinate system is chosen such that $x$ is the parallel and $y$ the
    transversal coordinate with respect to the waveguide.}
 \end{figure}

We consider the same "frying-pan" setup as in Refs.~\cite{B+92,PSS01}, i.~e.\ 
a cavity with an attached waveguide as shown in Fig.~\ref{sketch}. The wave
function must vanish on the boundary of the system. To obtain the
corresponding closed system, Dirichlet boundary conditions are imposed also on
the line $x=0$. It is assumed that this point of separation is well inside the
waveguide such that for $-a<x<\infty$ the wavefunction can be expanded in
transversal modes
\begin{equation}
\phi_{\lambda}(y)=\sqrt{2/b}\sin (\lambda \pi y/b)\,.
\end{equation}
A condition on the minimum value of $a$ will be given later. 

%%%%%%%%%%%%%%%%%%%%%%%%%%%%%%%%%%%%%%%%%%%%%%%%%%%%%%%%%%%%%%%%%%%%%%
%%%%%%%%%%%%%%%%%%%%%%%%%%%%%%%%%%%%%%%%%%%%%%%%%%%%%%%%%%%%%%%%%%%%%%

Energy eigenfunctions in a billiard satisfy the Helmholtz equation
$(\Delta+k^2)\psi({\bf r})=0$ with $k=\sqrt{2mE/\hbar^2}$ and ${\bf
  r}=\{x,y\}$. In particular, the scattering states $\psi_{\lambda}$ are in
the channel region given in terms of the S-matrix as
\begin{eqnarray}\label{sstate} 
\psi_{\lambda}({\bf r},k)&=&
\sum_{\lambda'}{\phi_{\lambda'}(y)\/\sqrt{k_{\lambda'}}}
\(\delta_{\lambda',\lambda}\e^{-\i k_{\lambda'}x}+
S_{\lambda',\lambda}\e^{+\i k_{\lambda'}x}\)\,,
\end{eqnarray}
where $k_{\lambda}=\sqrt{k^2-(\lambda\pi/b)^2}$ is the longitudinal momentum.
There is a finite number $\Lambda=[kb/\pi]$ of open modes with real
$k_{\lambda}$. All other modes are evanescent and give contributions to the
scattering states which decay exponentially into the waveguide. It is known
that their influence is negligible unless the energy nearly coincides with a
threshold for the opening of a new channel \cite{SS95,AR01}. For simplicity we
will disregard evanescent modes here. Then the $\Lambda\times \Lambda$
S-matrix is unitary, and therefore it is related to a Hermitian {\em reaction
  matrix} \footnote{Despite some minor differences in the definition we use
  this term here because of the strong analogy to \protect\cite{WE47}.} $K$ by
\begin{eqnarray}\label{SK}
K&=&-\i(I+S)^{-1}(I-S)\\
S&=&(I+\i\,K)^{-1}(I-\i\,K)\,.
\end{eqnarray}
As billiards are invariant under time-reversal, $S$ and $K$ are symmetric
matrices.

On the other hand we consider the eigenfunctions of the closed billiard which
vanish at $x=0$ and can therefore be represented as 
\begin{eqnarray}\label{billefD} 
\Psi_{n}({\bf r})
&=& \sum_{\lambda}{u_{n,\lambda}
\,\phi_{\lambda}(y)\sin(k_{n,\lambda}\,x)/k_{n,\lambda}}
\qquad(-a\le x<0)\,.
\end{eqnarray}
The corresponding eigenvalue is $k_{n}$, and $k_{n,\lambda}$ follows in
analogy to $k_{\lambda}$.  Again because of time-reversal symmetry the
coefficients $u_{n,\lambda}$ can be chosen real.  These coefficients are
obtained by projecting the normal derivative of the eigenfunction at the
interface on the transversal modes
\begin{equation}\label{nderiv}
u_{n,\lambda}=\int_{0}^{b}\d y\,\phi_{\lambda}(y)
\[{\d/\d x}\,\Psi_{n}({\bf r})\]_{x=0}\,.
\end{equation}
It is our goal to represent at arbitrary wave number $k$ the reaction matrix
$K$ (and thus via Eq.~(\ref{SK}) also the S-matrix) in terms of the discrete
set $\{k_{n},u_{n,\lambda}\}$. For this purpose we consider the Green's function
for the closed billiard which is defined by $(\Delta+k^{2})\,G({\bf r};{\bf
  r'},k)=\delta({\bf r}-{\bf r'})$, as this quantity can at the same time
be represented in terms of the S-matrix and by a spectral sum. Indeed we have
\cite{PSS01}
\begin{eqnarray}\label{GFdeco}
G({\bf r};{\bf r'};k)&=&
\sum_{n=1}^{\infty} {\Psi_{n}({\bf r'})\Psi_{n}({\bf r})\over k^{2}-k_{n}^{2}}
\\
\label{GFdeco2}&=&
\sum_{\lambda,\lambda'} \s_{\lambda}({\bf r},k)
\,g_{\lambda\lambda'}(k)\,\psi_{\lambda'}({\bf r'},k)
\qquad(-a\le x'\le x<0)\,.
\end{eqnarray}
with 
\begin{equation}\label{PsiND} 
\s_{\lambda}({\bf r},k)={\phi_{\lambda}(y)
\sin(k_{\lambda}x)/ \sqrt{k_{\lambda}}}
\end{equation}
and 
\begin{equation}\label{g}
g(k)=(I+S(k))^{-1}\,.
\end{equation}
The S-matrix of the scattering system is implicitly contained in
Eq.~(\ref{GFdeco2}), both in the scattering states $\psi_{\lambda}$ and in the
matrix $g(k)$. We can extract this information from the normal derivative of
Green's function at the interface, projected to transversal modes. After
inserting Eqs.~(\ref{sstate}) and (\ref{PsiND}) into Eq.~(\ref{GFdeco2}) we
have
\begin{eqnarray}\label{ddG}
\lim_{x'\to -0}\lim_{x\to -0}
{\partial^{2}\over
\partial_{x}\partial_{x'}}G_{\lambda,\lambda'}(x,x')
= 
\i\sqrt{k_{\lambda}k_{\lambda'}}\sum_{\lambda''}g_{\lambda\lambda''}
\(-\delta_{\lambda'',\lambda'}+
S_{\lambda'',\lambda'}\)
\label{g2x}
\end{eqnarray}
Using Eqs.~(\ref{g}) and (\ref{SK}) we see that the r.h.s. reduces 
essentially to an element of the reaction matrix 
\begin{equation}\label{KD}
K_{\lambda\lambda'}={1\/\sqrt{k_{\lambda}k_{\lambda'}}}\,
\lim_{x'\to -0}\lim_{x\to -0}
{\partial^{2}\over
\partial_{x}\partial_{x'}}G_{\lambda,\lambda'}(x,x')\,.
\end{equation}
Because of the symmetry of Green's function with respect to its spatial
arguments the result does not depend on the order of limits, just the
singularity occurring for $x=x'$ must be avoided.  Comparing Eq.~(\ref{KD}) to
the spectral decomposition of Green's function Eq.~(\ref{GFdeco}) one
naively expects that the reaction matrix has a spectral representation in the
form
\begin{equation}\label{KW}
K_{\lambda\lambda'}=\pi\,\sum_{n=1}^{\infty}\,
{W_{n,\lambda}\,W_{n,\lambda'}\/k^{2}-k_{n}^{2}}
\end{equation}
with coupling constants
\begin{equation}\label{oldW}
\tilde W_{n,\lambda}=u_{n,\lambda}/\sqrt{\pi k_{\lambda}}\,. 
\end{equation}
However, this is not the case, as it was found in Refs.~\cite{PSS01,SSS02}
that with Eq.~(\ref{oldW}) the sum (\ref{KW}) fails to converge. Convergence
is lost because the individual terms in the (absolutely convergent) series
Eq.~(\ref{GFdeco}) grow too much upon the differentiation required by
Eq.~(\ref{KD}). We will show in the following that Eq.~(\ref{KW}) is still
valid, albeit with modified coupling constants
\begin{equation}\label{W}
W_{n,\lambda}=
\sqrt{k_{\lambda}\/\pi}
{\sin\({k_{n,\lambda}\/k_{\lambda}}{\pi\/ 2}\)\over k_{n,\lambda}}
u_{n,\lambda}\,.
\end{equation}
Keeping in mind that differentiation and summation do not commute in
Eq.~(\ref{GFdeco}), we need a different strategy which allows to make use of
this series anyway: We will represent the second derivative of Green's
function in Eq.~(\ref{KD}) by the {\em value} of Green's function at some
shifted point in space. To implement this idea we return to Eq.~(\ref{g2x})
and bring it by some formal manipulations into the shape of
Eq.~(\ref{GFdeco2}), projected onto the transversal modes.  Then we get a
representation of the reaction matrix which is equivalent to Eq.~(\ref{KD})
but does not require differentiation, namely
\begin{eqnarray}\label{KG}
K_{\lambda\lambda'}
&=&
\i\,\sum_{\lambda''}g_{\lambda\lambda''}
\(-\delta_{\lambda'',\lambda'}+
S_{\lambda'',\lambda'}\)
\nonumber\\&=&
\sqrt{k_{\lambda}k_{\lambda'}}
\sum_{\lambda''}{\sin(-\pi/2)\over \sqrt{k_{\lambda'}}}
\, g_{\lambda\lambda''}
{1\over \sqrt{k_{\lambda''}}}\(\delta_{\lambda'',\lambda'}\e^{+\i\pi/2}+
S_{\lambda'',\lambda'}\e^{-\i\pi/2}\)
\nonumber\\&=&
\sqrt{k_{\lambda}k_{\lambda'}}\,
G_{\lambda,\lambda'}\(-{\pi/2 k_{\lambda}},-{\pi/2 k_{\lambda'}}\)\,.
\end{eqnarray}
Now we can safely use the spectral decomposition of Green's function
Eq.~(\ref{GFdeco}) together with Eq.~(\ref{billefD}) to obtain
\begin{eqnarray}\label{kmat-result}
K_{\lambda\lambda'}&=&\sqrt{k_{\lambda}k_{\lambda'}}\,\sum_{n}
{1\over k^2-k_{n}^2}
\,\sin\({k_{n,\lambda}\/ k_{\lambda}}{\pi\/2}\)
{u_{n,\lambda}\/k_{n,\lambda}}
\sin\({k_{n,\lambda'}\/ k_{\lambda'}}{\pi\/2}\){u_{n,\lambda'}\/k_{n,\lambda'}}\,.
\end{eqnarray}
This is indeed equivalent to Eq.~(\ref{KW}) with the coupling constants given
in Eq.~(\ref{W}). 

Some comments are in order at this stage.  The point
$x_{\lambda}=-{\pi/2k_{\lambda}}$ in Eq.~(\ref{KG}) corresponds to the first
maximum of the partial wave $\lambda$. Although no differentiation was
involved, the normal derivatives of the wave functions $u_{n,\lambda}$ appear
in Eq.~(\ref{kmat-result}) because they determine the amplitudes of the
partial waves at these maxima according to Eq.~(\ref{billefD}). However, the
full reaction matrix $K$ can be obtained by the outlined procedure only, if
the points $x_{\lambda}$ are inside the waveguide for all
$\lambda=1,\dots\Lambda$. Otherwise Eq.~(\ref{KG}) breaks down. This implies a
restriction which is at its strongest for the minimum of $k_{\lambda}$ at
$\lambda=\Lambda$. We can represent the wave number as
$k=(\Lambda+\kappa)\pi/b$ with $0\le\kappa<1$ and find
$k_{\Lambda}=\sqrt{2\Lambda\kappa+\kappa^2}{\pi/b}$. Hence, our expression for
the reaction matrix will be valid provided that
\begin{equation}\label{condition}
{b/2a}\le \sqrt{2\Lambda\kappa+\kappa^2}\,.
\end{equation}
This condition will always be violated at the threshold energies for the
opening of new scattering channels where $\kappa=0$. On the other hand, if we
avoid these singular points by fixing $\kappa$ to some positive value, the
condition will always be satisfied in the semiclassical limit
$\Lambda\to\infty$. In some sense these restrictions on the validity of our
approach are similar to those allowing to neglect evanescent modes
\cite{DS92a,SS95,AR01}. From Eq.~(\ref{condition}) we conclude that the
situation which is most favourable for reaction-matrix theory in Dirichlet
billiards is a closed billiard which extends very far into the waveguide $b\ll
a$. Intuitively this should be clear, because then the shape of the billiard
resembles a scattering system. 
 
A second comment concerns the uniqueness of the suggested procedure.
According to Eq.~(\ref{GFdeco2}), and because of the special geometry we
consider, any point $-a\le x\le\infty$ inside the waveguide region can be used
to extract information about the S-matrix, while we have selected only those
points where some partial wave $\lambda$ has maximum amplitude. This may
appear like an arbitrary choice. Indeed, if we sample Green's function at
sufficiently many points, arbitrarily placed inside the waveguide, it is in
principle possible to compute the S-matrix or the reaction matrix.  However,
the resulting expression will not have the canonical form of Eq.~(\ref{KW}) and
no coupling constants for individual levels can be defined in this case.
Therefore the connection to the formalism of Refs.~\cite{WE47,Feshbach,MW69} would
be lost within such an approach and consequently it would not be very useful.
In other words, the main accomplishment of our theory is not the mere
possibility to compute the S-matrix from data obtained in the closed billiard,
rather it is the fact that the resulting expression is still of the form of
Eq.~(\ref{KW}).

Further we note that the coupling constants in Eq.~(\ref{W}) reduce to the
naively anticipated Eq.~(\ref{oldW}) for states $n$ which are close to the
energy shell of scattering, i.~e.\ $W_{n,l}\sim u_{n,l}/\sqrt{k_{\lambda}\pi}$
for $k_{n}\sim k$. This explains why Eq.~(\ref{oldW}) works (up to some degree
of accuracy) for a cluster of almost degenerate states \cite{SSS02}, and why
it also provides a consistent answer for the mean coupling strength in the
semiclassical regime \cite{PSS01}: The approximation $k_{n}\sim k$ was always
made from the outset. In general this approximation is not justified and
Eq.~(\ref{oldW}) breaks down.

The crucial difference between Eqs.~(\ref{oldW}) and (\ref{W}) is their
behavior for $n\to\infty$ which decides the issue of convergence. To discuss
this question we need the semiclassical estimate for the average behavior of the
boundary functions ${\langle|u_{n,\lambda}|^{2}\rangle/k_{n,\lambda}}\sim
{4/A}$ which we derived in Eqs. (34), (38) of Ref.~\cite{PSS01} for chaotic
billiards (see \cite{B+02} for more general and accurate estimates of this
type). $A$ denotes here the area of the billiard. Substitution into
Eq.~(\ref{W}) leads to
\begin{equation}\label{asymp}
{|W_{n,\lambda}W_{n,\lambda'}|\over k^2-k_{n}^{2}}\sim k_{n}^{-3}\sim n^{-3/2} 
\qquad(n\to\infty)\,,
\end{equation}
i.~e.\ Eq.~(\ref{KW}) is indeed absolutely convergent.

%%%%%%%%%%%%%%%%%%%%%%%%%%%%%%%%%%%%%%%%%%%%%%%%%%%%%%%%%%%%%%%%%%%%%%%%%%%%%%%

We find it illuminating to apply presented theory to a simple toy model, where
all relevant quantities are known in closed form.  Consider a rectangular
billiard with side lengths $a$ and $b$ which yields a half-infinite empty
waveguide of width $b$ if one of the walls is removed.  This is an integrable
system since the transversal modes of the waveguide are decoupled. Therefore
we can restrict attention to some particular mode $\lambda$. The corresponding
diagonal element of the S-matrix is found to be
\begin{equation}\label{srb}
S_{\lambda}=-\e^{2\i k_{\lambda}a}\,.
\end{equation}
This is to be reproduced by the spectral sum Eq.~(\ref{KW}). The normalized
eigenfunctions of the closed billiard are
\begin{equation}
\Psi_{\mu,\lambda}(x,y)={2/ \sqrt{ab}}\sin(\mu\pi x/a)\,\sin(\lambda\pi y/b)
\end{equation}
and therefore we have the longitudinal momentum $k_{\mu,\lambda}=\mu\pi/a$, 
the boundary function $u_{\mu,\lambda}=\sqrt{2/a}\,k_{\mu}$
and the coupling constants
\begin{equation}
W_{\mu,\lambda}=\sqrt{2 k_{\lambda}\/a \pi}
{\sin\({\pi^2\/ 2a}{\mu\/k_{\lambda}}\)}\,.
\end{equation}
Substitution into Eq.~(\ref{KW}) yields after some straightforward
transformations
\begin{eqnarray}\label{kseries}
K_{\lambda}&=&{ak_{\lambda}\/\pi^2}\sum_{\mu=1}^{\infty}
{1-\cos({\pi^2\/a}{\mu\/k_{\lambda}})\over (ak_{\lambda}/\pi)^{2}-\mu^{2}}\,.
\end{eqnarray}
According to the two terms in the numerator we split this expression into two
series which can separately be evaluated with the help of Eq.~(1.445.6) from
Ref.~\cite{Gradstein}. Upon recombination of the two results we
have
\begin{eqnarray}\label{krb}
K_{\lambda}&=&\cot{k_{\lambda}a}\,,
\end{eqnarray}
which is precisely the reaction matrix corresponding to the S-matrix
Eq.~(\ref{srb}) via Eq.~(\ref{SK}).

%%%%%%%%%%%%%%%%%%%%%%%%%%%%%%%%%%%%%%%%%%%%%%%%%%%%%%%%%%%%%%%%%%%%%%
 \begin{figure}[tb]
  \centerline{
    \psfig{figure=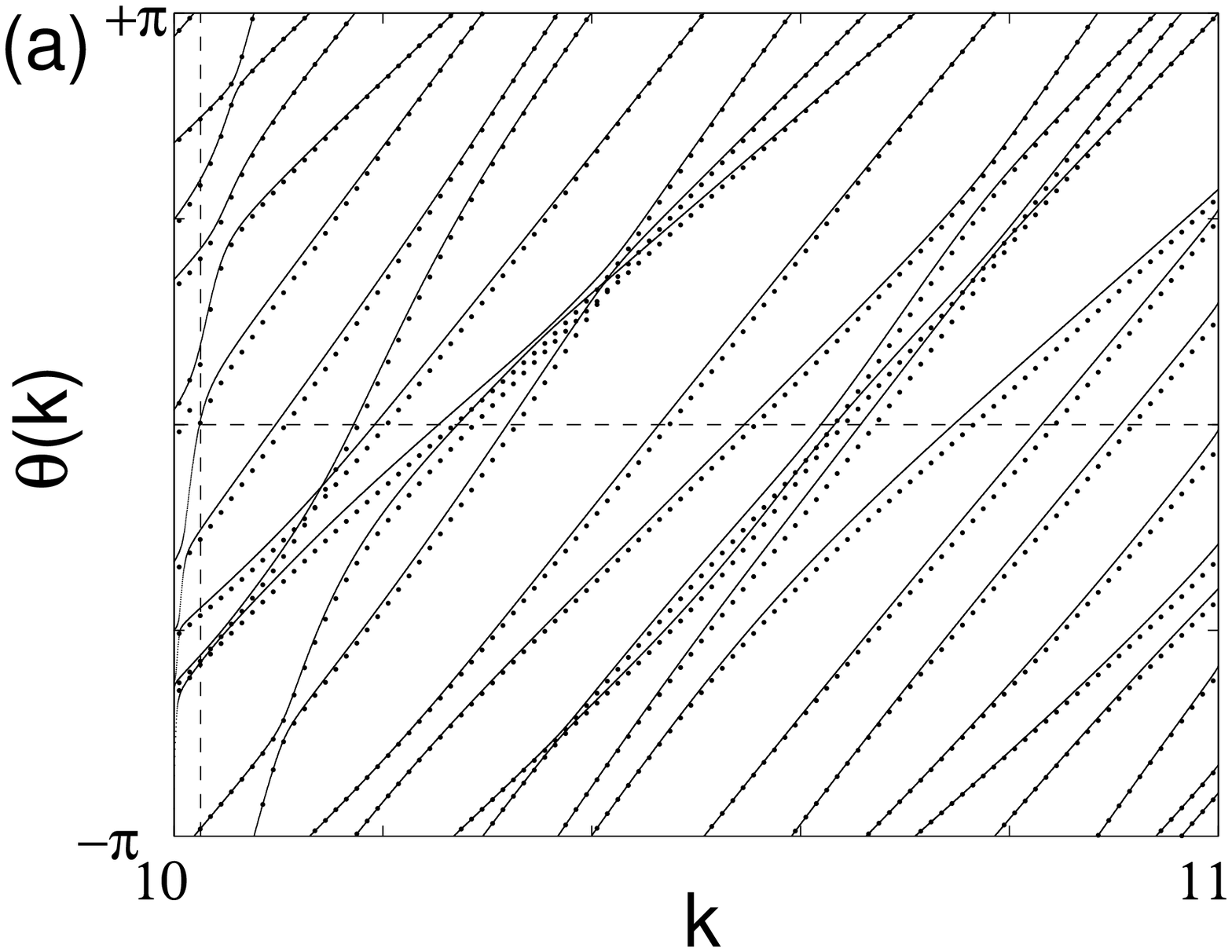,width=80mm}
    \psfig{figure=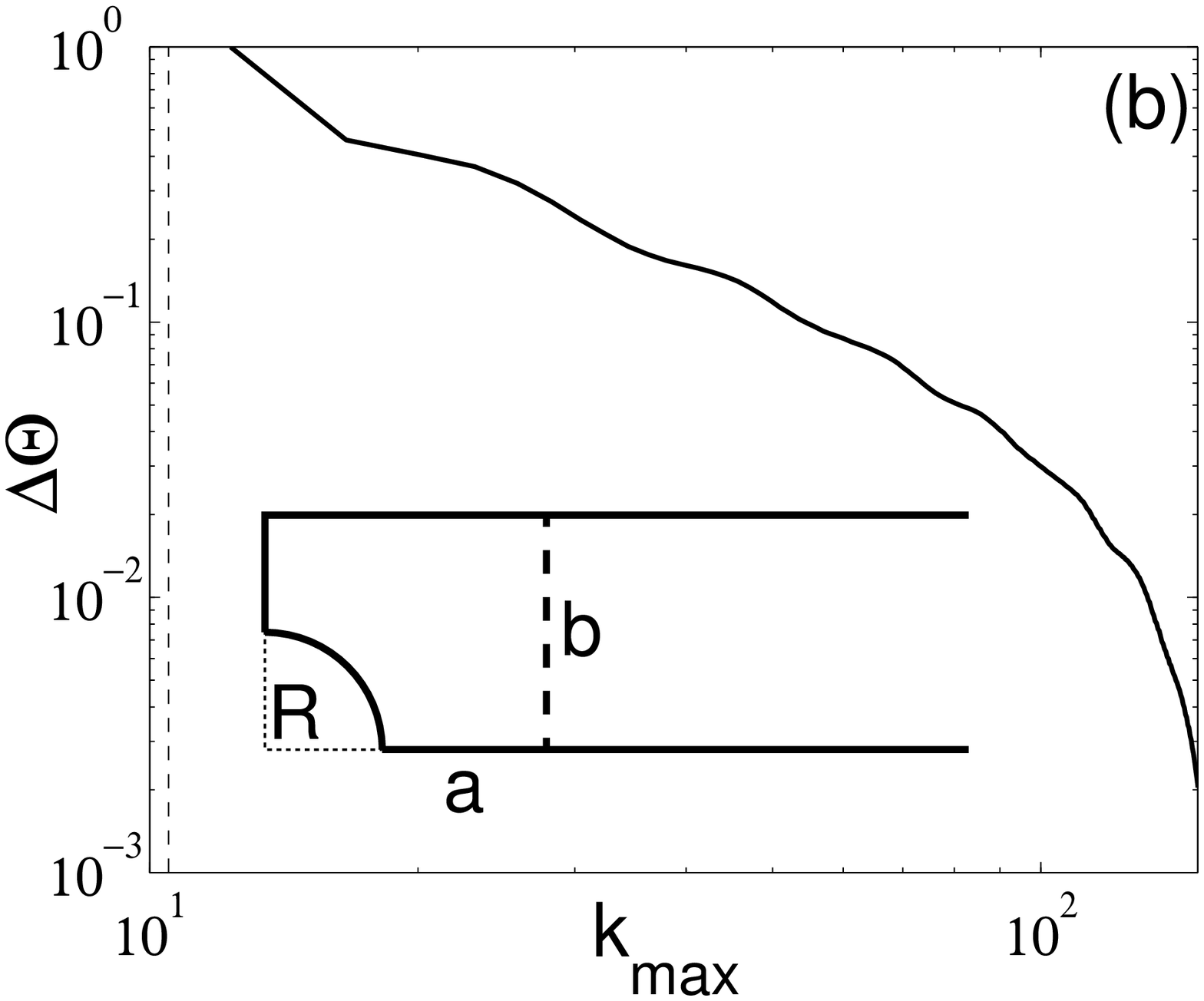,width=80mm}
  }
  \caption{\label{sinai}
    In the energy interval with 10 open modes the numerically exact S-matrix
    is compared to the result of Eq.~(\protect\ref{KW}) for the quarter Sinai
    billiard shown in the inset ($b=\pi$, $R=\pi/2$, $R+a=5\pi/4$).  (a) Shown
    are the eigenphases $\theta_{\lambda}(k)$ ($\lambda=1,\dots 10$) as a
    function of the wave number $k$. Full line: exact result. Dots:
    approximation based on all 1,000 eigenfunctions up to $k_{\rm max}=36.27$.
    Vertical line: Range of validity of Eq.~(\protect\ref{KW}) as given by
    Eq.~(\protect\ref{condition}).  (b) Shown is the deviation of the total
    phase $\Theta=\sum_{\lambda}\theta_\lambda$ from the exact result for
    some fixed wave number $k=10.50$ (vertical line) but with varying cutoff $k_{\rm
      max}$ for the included billiard eigenfunctions.}
 \end{figure}
%%%%%%%%%%%%%%%%%%%%%%%%%%%%%%%%%%%%%%%%%%%%%%%%%%%%%%%%%%%%%%%%%%%%%%
 
 Now we test the outcome of Eq.~(\ref{KW}) in the Sinai billiard.  S-matrix
 and eigenstates were computed with the methods described in Ref.~\cite{SS95}.
 For Fig.~\ref{sinai} we have chosen a large number of open modes,
 $\Lambda=10$, and a geometry which does not allow for narrow resonances. This
 is a very demanding regime, as a high number of states is expected to
 contribute. Nevertheless, in Fig.~\ref{sinai}a we show that a reasonable
 agreement can already be obtained from the 1,000 lowest internal modes. The
 agreement improves as we increase the number of states (Fig.~\ref{sinai}b).
 For $k_{\rm max}=150$, corresponding to 15,575 modes, hardly any deviation is
 visible on the scale of Fig.~\ref{sinai}a (not shown). In contrast, from
 Eq.~(\ref{oldW}) we get garbage irrespective of $k_{\rm max}$ (also not
 shown).  Note that the error in an individual eigenphase $\theta_\lambda$ of
 the S-matrix is at its largest when $\theta_\lambda=0$.  These are the
 eigenvalues for a billiard with identical geometry but Neumann b.~c.~at
 $x=0$, while at $\theta_\lambda=\pm\pi$ the Dirichlet billiard is quantized
 and we get perfect agreement. Had we chosen Neumann b.~c.~as starting point
 for the evaluation of the reaction matrix, the result would show relatively
 large deviations close to $\theta_\lambda=\pm\pi$. In this sense the
 different boundary conditions are complementary. Finally we remark that from
 Eq.~(\ref{condition}) we cannot expect agreement in the interval $10\le k\le
 10.025$ in Fig.~\ref{sinai}a.  Indeed we observe in this region a few points
 which are very far off the exact result.

Summarizing we have shown how and under what circumstances the reaction matrix
of a cavity with attached waveguides can be represented as a spectral sum of a
closed billiard with the physically relevant Dirichlet boundary conditions.

\paragraph*{Acknowledgements}
Stimulating discussions with D.~V. Savin, R.~Schubert and V.~V.~Sokolov are
gratefully acknowledged.

%%%%%%%%%%%%%%%%%%%%%%%%%%%%%%%%%%%%%%%%%%%%%%%%%%%%%%%%%%%%%%%%%%%%%%%%%%%%%%%

\end{document}